# Barriers to grid-connected battery systems: Evidence from the Spanish electricity market

Yu Hu,[a,*] David Soler Soneira,[a] María Jesús Sánchez,[b]

[a] *Invesyde S.L., Madrid, Spain*

[b] *Escuela Técnica Superior Ingenieros Industriales, Universidad Politecnica de Madrid, Spain*

**Abstract**

Electrical energy storage is considered essential for the future energy systems. Among all the energy storage technologies, battery systems may provide flexibility to the power grid in a more distributed and decentralized way. In countries with deregulated electricity markets, grid-connected battery systems should be operated under the specific market design of the country. In this work, using the Spanish electricity market as an example, the barriers to grid-connected battery systems are investigated using utilization analysis. The concept of "potentially profitable utilization time" is proposed and introduced to identify and evaluate future potential grid applications for battery systems. The numerical and empirical analysis suggests that the high cycle cost for battery systems is still the main barrier for grid-connected battery systems. In Spain, for energy arbitrage within the day-ahead market, it is required that the battery wear cost decreases to 15 €/MWh to make the potentially profitable utilization rate higher than 20%. Nevertheless, the potentially profitable utilization of batteries is much higher in the applications when higher flexibility is demanded. The minimum required battery wear cost corresponding to 20% potentially profitable utilization time increases to 35 €/MWh for energy arbitrage within the day-ahead market and ancillary services, and 50 €/MWh for upward secondary reserve. The results of this study contribute to the awareness of battery storage technology and its flexibility in grid applications. The findings also have significant implications for policy makers and market operators interested in promoting grid-connected battery storage under a deregulated power market.

---

[*] Corresponding author. Tel: +34-917395257; Email: yu.hu@invesyde.com

# 1. Introduction

Electrical energy storage is considered to be essential for the future energy system with an increasing penetration level of intermittent energy sources [1]. The battery storage industry has been growing rapidly during the last few years, battery systems are becoming more attractive with falling costs and improving performance [2]. Currently, the number for grid-connected electrochemical storage systems has reached 1600 MW in terms of power capacity in 2017 (almost three times higher than in 2014), where Li-ion batteries account for about 81% (1300 MW) of the total electrochemical capacity. [3]

However, as an emerging technology, grid-connected battery systems face a number of barriers which limit their further deployment. According to [4], despite the significant reduction, the initial cost for grid-connected battery storage projects is still high for investors. At the same time, the lack of awareness of the technological and economic benefits of such systems keeps the shareholders from adapting to the new asset class.

One of the problems of a grid-connected battery storage project is the market issues. Nowadays, although the structure of electricity markets/systems in each country vary in terms of regulation and level of competition, a general trend toward deregulated and integrated markets has been observed globally in the recent decades [5] [6] [7]. In countries with deregulated electricity markets, the feasibility assessment and practical operation of such battery systems should be conducted under its specific market and regulatory framework.

The goal of this work is to develop a framework for analyzing the barriers and identifying future potential applications focused on grid-connected battery systems under deregulated electricity markets. In this paper, utilization analysis has been applied taking the Spanish electrical power system as an example. The concept of "potentially profitable utilization time" is proposed and applied to different market applications. At the same time, the potential changes in utilization time emerging from the battery cost reduction are also investigated, which is crucial for evaluating future scenarios.

The remaining of this paper is structured in the following way. Section 2 provides a literature review related to the topic and lists the main contributions of this paper. Section 3 introduces the market designs of both wholesale markets and ancillary services in Spain and discusses the potential application of battery storage systems. Section 4 presents the proposed utilization analysis for different market applications. Finally, section 5 discusses the other related issues and sets out the main conclusions of this paper.

# 2. Literature review

Grid-connected battery storage systems have shown a great potential in power grid applications. For example, battery storage can be applied in power system regulation by providing frequency control, power reserves and balancing power [4]. When combined with renewable energy sources such as wind power, a battery system may store the excess power generation when necessary and stabilize the power output of renewable energy generators [8]. At the same time, batteries can also increase power system reliability by providing peak load and resolving congestion. Compared with other technologies, battery systems may provide system flexibility in a more distributed and decentralized way [9].

The cost issue remains the most significant obstacle for grid-connected batteries. Despite the fact that battery energy storage technologies, especially the Li-ion battery, have experienced significant cost reductions, the cost for grid-connected battery storage projects is still high compared with mature technologies such as pumped hydroelectric storage [1] and [10].

The key factor that determines the cost barrier to the current battery technologies for grid scale applications is their lifetime. All batteries have a finite life since every charge-discharge cycle results in some degradation [11]. Research works have shown that battery degradation rates is higher during the early cycles than the later cycles, and finally the degradation rates would increase again when the battery reaches its end of life [12].

The general lifetime of battery systems can be described as cycle life which is defined as the number of complete charge–discharge cycles that the battery can perform before its nominal capacity falls below 80% of its initial rated capacity [13]. Besides capacity fading, other effects of battery degradation such as impedance rising may also cause problems for practical applications [14].

The cost of battery degradation can be defined as battery wear cost, which is cost of the delivered energy from the battery. The battery wear cost can be calculated as the battery cost in €/kWh divided by the Equivalent Full Cycles until the battery is replaced [14]. It can be also considered as the depreciation the battery storage by the energy it could deliver throughout its lifetime. Zubi et al. stated in 2018 that Li-ion batteries were still very far from the cost competitive range for grid-connected use given the battery specific cost around 300/kWh with a cycle life around 2000 cycles [15]. And for real battery storage products, Telsa provides a warranty that guarantees 70% energy retention in 10 years with a maximum number of 2800 Equivalent Full Cycles for its Li-ion battery product "Powerwall" [16].

Given the numbers mentioned above, for an empirical battery storage with an initial cost of 300 €/kWh and a conservative estimation of totally 3000 equivalent full cycles lifetime before the battery is replaced. It would imply a pure battery wear cost of 0.1 €/kWh (100 €/MWh) if we considered that the total initial cost will be equally amortized by the total energy delivered from the battery through its lifetime.

It should be noted that, besides Li-ion batteries, other battery technologies such as Vanadium flow, NaS, and advance Lead acid may become attractive in the future given that future cost reduction and higher cycle life are achieved [17]. The major drawback of these technologies such as low energy density, safety and maintaining requirements are not considered to be significant barriers for grid-scale applications.

Apart from technical issues, research works related to grid-connected battery systems have been conducted on feasibility studies and economic assessments using different optimization models. In most of the cases, energy storage systems are studied as an internal component of a micro-grid [18] [19] [20] or a hybrid system [21] [22] [23], where the operational strategy of energy storage depends not only on the market, but significantly on the other resources/loads within the system. The main research objectives of the approaches mentioned above were not focused on the energy storage and scarce information is presented related to the interaction between the energy storage component and the electricity market.

Indeed, compared to the traditional technologies, the cost of battery storage is still much higher for grid-scale applications. Many recent studies have been conducted for end-user applications [24] [25], isolated power grid [26] and local markets [27], where the energy price is higher than the grid applications in most of the electricity markets.

Meanwhile, a general under-estimation of battery cost has been observed in many studies when the electricity markets were involved. For example, in [21], a 10 DKK/MWh (1.34 €/MWh) battery degradation cost was used in the numerical study, while no initial installation cost has been considered. And in [23], the cost of usage for a battery storage system is estimated as 7.22 €/MWh.

Among other research works related to stand-alone battery systems, Mirtaheri et al. [28] studied the optimal planning and scheduling of battery storage in distribution networks. However, this work only analyzed the battery system using simulated low voltage local distribution network.

Olk et al. [29] investigated the bidding strategy for batteries in the German secondary reserve market, and concluded that stand-alone battery storage is not profitable currently in the secondary reserve market. Research works have also been conducted to evaluate the potential participation for battery storage in other commonly applied ancillary services such as primary regulation [30] [31], solving transmission and distribution constraints [32] [33], and peak-load shaving [34].

Even so, the focuses of the works mentioned above are still narrow and the results are still limited due to the simplification of market conditions [32] [33] and pool price modeling [29], and the limitation in economic benefit analysis [30] [31] [34]. Moreover, the feasibility studies and economic analyses mentioned above did not provide a general picture or sensibility analysis of the potential profitability of batteries for grid applications.

In order to shape a future pathway of grid-connected batteries technology, a utilization analysis is conducted in this paper. Utilization analysis [35] is a common method for economical assessment of renewable energy projects. Concepts such as "Equipment Utilization Hours [36]" and "Degree of Utilization [35]" are widely used to measure and compare revenue from wind, solar and hydropower.

Rather than measuring by total energy delivered from the device, as most utilization analyses have done, the utilization in this paper is measured as potentially profitable utilization time. Compared to other measurements, the utilization factor defined by potentially profitable utilization time may better reveal the market potential for battery systems due to the following reasons.

Firstly, since the battery cost is still high, for most grid applications, economic valuation methods, such as net present value, or internal rate of return, will result in negative values. Moreover, from a more general point of view, the potentially profitable utilization time would be a more effective measurement for many stakeholders such as system operators, policy makers and technology promoters.

Secondly, in many electricity markets, the remuneration is calculated not only in terms of energy, but also power capacity. Measuring only the energy delivered by the battery would disregard the capacity payment.

Lastly, potentially profitable utilization time is a valuable measurement to compare the market potential for different applications with different remuneration structure.

This paper contributes to the literature by applying utilization analysis to assess the market performance of stand-alone battery systems in grid applications under deregulated electricity markets. The analysis is conducted under a real-world market design. Moreover, by addressing the potential declining of battery cost as future scenarios, the proposed method may provide a general overview of possible future market potential for battery systems. By comparing the sensibility analysis of potentially profitable utilization time for different applications, it highlights the competitive advantage of batteries in some specific applications. Thus, it bridges the gap between the still high, but falling, cost of battery storage technology and the identification of potential promising application in deregulated electricity markets.

## 3. Electricity market and ancillary services in Spain

In Spain, electrical utilities trade their energy production or power capacity through the wholesale electricity market or by providing ancillary services. Table 1 presents the electricity markets, ancillary services and their products in Spain.

### 3.1. Wholesale electricity market

OMIE (Operador del Mercado Ibérico de energía - polo Español) runs the wholesale electricity market in both Spain and Portugal. The market is composed of a set of sub-markets in which power generators and consumers trade hourly energy production products.

In general, the commonly referred to as Spanish electricity price is the market clearing price of the day-ahead market opened at noon for the next day with 24 hourly products. Power purchase and sale offers are matched according to their economic merit order, [37]. The settlement price of energy over a specific hour is determined by the point at which the supply and demand curves intersect, according to the marginal pricing model adopted by the European Union.

Table 1. Wholesale electricity market and ancillary services in Spain.

|  | Market/Service | Product |
|---|---|---|
| Wholesale Market<br>*Operated by OMIE* | Day-ahead market | Energy |
|  | Intraday market |  |
| Ancillary services<br>*Operated by REE* | Primary reserve | (Non-remunerable) |
|  | Secondary reserve | Capacity and Energy |
|  | Tertiary reserve | Energy |
|  | Deviation management |  |
|  | Management of Technical constraints |  |
|  | Interruptibility Service | Capacity and Energy |
|  | Voltage control | Reactive power |

After the day-ahead market, market participants may adjust their position through intraday markets. There are in total six intraday markets which are held several hours earlier than the delivery of power production [38]. Again, the settlement prices are determined by the power supply and demand curves. There is also a continuous intraday market which runs in parallel where real-time bid and asks are matched like the common exchange market.

Similar to other energy storage technologies, battery storage systems are able to perform energy arbitrage by buying energy when the energy price is low and selling the stored energy at the peak hours with high energy prices.

### 3.2. Ancillary services

Ancillary services are defined as the "set of products separated from the energy production, which are related to security and reliability of a power system" [39]. In Spain, REE (Red Eléctrica de España) operates the ancillary services market.

#### 3.2.1. Primary reserve

The Spanish power grid is a part of the Union for the Coordination of the Transmission of Electricity (UCTE) power system. According to the handbook of UCTE [40], "the objective of primary control is to maintain a balance between generation and consumption (demand) within the synchronous area. By the joint action of all interconnected parties / Transmission System Operators (TSO), primary control aims at the operational reliability of the power system of the synchronous area and stabilizes the system frequency at a stationary value after a disturbance or incident in the time-frame of seconds, but without restoring the system frequency and the power exchanges to their reference values."

Frequency Containment Reserves (FCR) is the main primary reserve service providing balancing capacity in many European countries, including Austria, Belgium, Netherlands, France, Germany and Switzerland, and it is expected that the Western Denmark transmission system operator will join the common FCR market in the future. The common FCR market operates with weekly-ahead auctions with weekly symmetric product.

Unlike the market-based services mentioned above, the primary control in Spain has been defined as a mandatory non-remunerable service: generating units must be capable of modifying 1.5% of their rated output in less than 15 s for frequency variations less than 100 mHz, and linearly up to 30 s for frequency deviations up to 200 mHz [39]. As a result, there is no profit for battery storages for participating in primary reserve in Spain.

### 3.2.2. Secondary reserve

The UCTE [40] defines secondary reserve as an ancillary service that "maintains a balance between generation and consumption (demand) within each control area as well as the system frequency within the synchronous area, taking into account the control program, without impairing the primary control that is operated in the synchronous area in parallel."

Automatic generation control (AGC) systems are used in each control area to adjust the active power output of generation units that are participating in the service. In Spain the AGC orders are sent every 4 seconds, and the secondary reserve should respond in 100 seconds when required and should be maintained for 15 minutes.

Secondary reserve is hired by the system operator using a specific day-ahead secondary reserve market. Regulating zones (major energy groups) submit bids of upward and downward power capacity reserve (in MW) of their generating units with associated price (in €/MW) of the capacity in an hourly basis. Then the system operator contacts the power capacity band on a least cost basis until the required capacity (calculated by the system operator) is reached. The verification and settlement of the secondary reserve is performed in level of regulating zones, penalties will be applied when a regulating zone does not comply with the response criteria.

Besides the income from power capacity band, the remuneration of the secondary reserve contains an energy term (in €/MWh). The energy deviation due to secondary control operation is priced at the substituting tertiary energy that would result if the associated tertiary reserve market were called [39]. At the same time, penalties will be applied when a control zone does not comply with the response criteria. Theoretically, battery storage systems are able to participate in this service and be remunerated in terms of power capacity and energy delivered.

### 3.2.3. Tertiary reserve and deviation management

The objective of tertiary reserve is to restore the used secondary regulation band. Previously authorized generating units (Normally conventional thermal power producers, pumped hydro storage systems and authorized manageable renewable energies) are forced by law to offer available power in the tertiary market [41]. The bids are sent at 23:00 of the day before the delivery day, and may be updated till 25 minutes before the beginning of the programming hour. Tertiary regulation is allocated and marginal upward/downward prices are determined by the system operator using economic order 15 minutes before the programming hour (if necessary, during the programming hour) [38]. Generating units are expected to respond within 15 minutes and be able to maintain the services for at least two consecutive hours.

Deviation management aims to solve foreseen power imbalances (for example, unavailability of generation units or justified changes communicated from generation) maintained during several hours. The system operator will call the service when it has foreseen a large power difference between programmed power production and demand after the intraday market [38]. Again, marginal upward/downward prices are determined by the system operator using economic order.

Due to the limited notification time, the volatility of the energy prices in these two services is higher than the price volatility in the day-ahead market or the intraday adjustment market. Battery storage systems may arbitrage in these two services for higher price volatility.

### 3.2.4. Management of technical constraints

The system operator performs the management of technical constraints after the day-ahead market, when the power generation and demand resulting from the day-ahead market does not comply with power transmission constraints or security criteria. During this process, the system operator holds auctions that aim to increase or decrease the scheduled power production of power plants and finally to solve the technical constraints.

Technical constraints are solved with a two phase auction process. The system operator starts with allocates new bids that may solve the constraints in the first phase, and then, in the second phase, bids that may re-establish the balance between power generation and demand. Unlike other markets where the marginal clearing price is used, both bidding processes are based on economic merit order and each participant is paid with the price associated with the bids. Currently the first auction phase of the service is only provided by large thermal power plants.

### 3.2.5. Interruptibility Service

Interruptibility service is a demand-side response program provided only by the large industrial consumers to reduce the power consumption according to the order issued by the system operator.

The objective of the service is to reduce the extraordinary and temporary shortage of power generation caused by peak demands or sudden decrease of renewable generation due to weather changes. In other EU countries, this service is normally classified as a strategic reserve.

The provider of the service receives a capacity payment. The allocation of the service is through a competitive allocation mechanism with a face-to-face bidding process managed by the system operator to ensure minimum costs. Currently, the service is standardized by two types of products with capacity blocks of 5 MW and 40 MW respectively, and the two products are auctioned separately. The provider of the service also receives financial compensations in terms of energy (MWh) at the price of tertiary reserve for the corresponding time period.

### 3.2.6. Voltage control service

Voltage control service requires power generators and qualified consumers who participate in this service generate reactive power and maintain the system voltage. In Spain this service is partially compulsory and non-remunerable, and partially remunerable based on performance evaluation. Due to the importance of voltage control to maintain the system security and reliability, the system operator defines a minimum mandatory proportion. The reactive capacity exceeding the minimum might be offered and remunerated at a fixed regulated price if the system operator accepts it [39].

## 3.3. Imbalance cost

In Spain, if the final production (consumption) of a normal power producer (consumer) is different from the final scheduled program, without resulting from the ancillary services, an imbalance cost will be applied. The imbalance cost depends on the general state of the grid system. If the grid system faces energy shortage, additional costs will be applied to the power producers (consumers) who produce less (consume more) energy than scheduled, while the power producers (consumers) who produce more (consume less) energy than scheduled will be considered helpful for the system and no additional imbalance cost will be charged. In the other case, if the grid system faces energy surplus, additional imbalance costs will be only charged to the power producers (consumers) who produce more (consume less) energy than scheduled. The additional cost will be calculated according to the usage of deviation management, tertiary reserve and secondary reserve in the corresponding time period.

A power producer is able to participate in the ancillary services after passing a set of tests from the system operator. If a power producer is participating in any ancillary services, the imbalance cost would not be applied to this generator. Still, other penalties would be applied if it does not fulfill the service requirements.

It is generally accepted that when working together with non-dispatchable power plants, such as renewable generators, batteries may help in reducing the imbalance cost by stabilizing the energy output.

## 3.4. Other services

Currently, there exists an ancillary service named additional upward reserve, aiming to ensure there is enough upward capacity to be offered in ancillary services. This service is provided by dispatchable thermal generators that have not been committed in the day-ahead market. When additional upward reserve is called by the system operator, these units may bid its available capacity, and receive the marginal price (in €/MWh) from the additional upward reserve market, if its bid is accepted. In exchange, thermal units that committed in the additional upward reserve market must sell their minimum capacity level to the intraday market, and submit bids in the ancillary services. However, this service is expected to be cancelled in the near future.

## 4. Utilization analysis

In this section, a utilization analysis is conducted through measuring the potentially profitable utilization time of grid-connected battery storage systems for common applications. We consider an empirical battery storage facility cost of 300 €/kWh and a conservative estimation of totally 3000 equivalent full cycles lifetime before the battery is replaced. And it is considered that the total cost will be equally amortized by the total energy delivered from the battery through its lifetime, each kWh (cycle) of energy from the battery system would imply a pure battery wear cost of 0.1 €/kWh (100 €/MWh).

In fact, the battery degradation has been simplified since the term equivalent full cycles is used. In other words, the total energy delivered from the battery will be equivalent to 3000 cycles for its whole lifetime. However, the total number of cycles in the real operation will be higher than 3000 since the real energy capacity of the battery will be less than the nominal capacity and decrease continuously throughout the lifetime due to the degradation.

It should be noted that the battery cost normally includes the cost of energy storage units (in €/MWh) and the cost of the power system (normally in €/MW). In practice, battery systems are normally designed with the energy capacity that ensures the battery to be operated at the nominal power capacity for a certain number of hours (1-hour, 2-hour, and 4-hour are the most commonly used battery design). In this analysis, for the sake of simplicity, it is assumed that the cost of the power system is included in the total cost of 300 €/MWh of the battery system.

In this analysis, it is considered to as "potentially profitable" when the remuneration of a certain operation is higher than the corresponding battery wear cost caused by delivering energy from the battery. Due to the fact that the battery cost is declining rapidly, a sensitivity analysis is conducted by investigating how changes in battery cost would affect the potentially profitable utilization time, which provides a general scenario for the future market potential.

It should be noticed that a general formula does not exist in measuring the potentially profitable utilization time, since the remuneration structure varies for each application or service. For example, the capacity payments may vary in time periods from minutes to months, and the energy payment may vary in prices based on the market condition. For each application, the payments in energy term and power capacity term should be calculated under its specific market design.

## 4.1. Energy arbitrage

In deregulated electricity markets, the price of electricity is determined by the power supply and demand, Figure 1 presents the boxplot of the hourly prices in the day-ahead market of Spain from 2015 to 2019 in a monthly level. The market price varies generally between 0 €/MWh and around 90 €/MWh. The main factors that affect the market price may include, but not limited to, power demand, renewable power generation, coal and gas prices, emission prices, and energy prices in the neighboring countries.

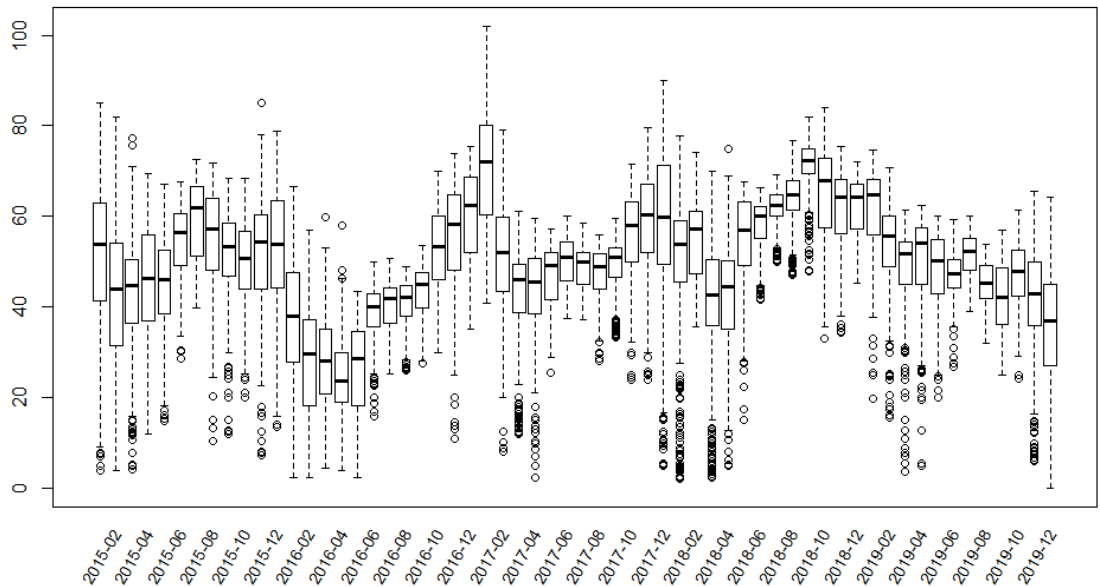

**Figure 1. Boxplot of the hourly prices in €/MWh in the day-ahead market of Spain from 2015 to 2019 in a monthly level.**

It also should be noted that in the Spanish electricity markets the prices associated in the bids and offers are limited to being between 0 €/MWh and 180.3 €/MWh, while in many other countries, the limitation for the highest price is much higher and negative prices are allowed.

In a seasonal/yearly level, the market prices are generally affected by the seasonal demand pattern, the seasonality of hydro power production, the fuel and $CO_2$ prices, and the interconnected electricity market. For example, in late 2016/early 2017, the market price in Spain reached its highest level since France shut down its nuclear power plants for inspection and the high demand in winter.

On the other hand, the main factors for the variation of the market price in a daily/weekly level are mainly the renewable energy production, and the daily/weekly pattern of power demand curve. Figure 2 presents the boxplot of the hourly prices in November 2019 in a daily and hourly basis.

Energy arbitrage is practiced by buying energy from the grid at a low price and selling it back at a higher price [42]. Currently in Spain, energy itself as a single product is traded in the wholesale market, and used to provide some of the ancillary services such as management of technical constraints, tertiary reserve and deviation management, meanwhile, battery storage can be installed together with renewable power generators, reducing their imbalance cost when the actual power generation is different from the scheduled generation and the direction of the imbalance is the same as total the system imbalance (in this case additional penalty will be applied as the imbalance cost).

At the same time, the potential profit of energy arbitrage also depends on the operation frequency. In this study, energy arbitrage using battery storage system is simulated in daily cycles. In other words, the battery system will perform one full charge-discharge cycle daily. The reasons of the daily operation cycle are as follows: Firstly, due to the demand difference between day and night, a daily pattern of electricity price with peak and valley hours is generally observed. Secondly, the daily operation aligns with the designed operation pattern for many battery producers in terms of designed lifetime and warranty provided by the manufacturer. And lastly, the most common energy-power capacity design for battery systems remains at the level of several hours. Thus, it would not be meaningful to perform energy arbitrage in longer time periods, such as monthly to yearly with a battery system which stores energy for only a few hours.

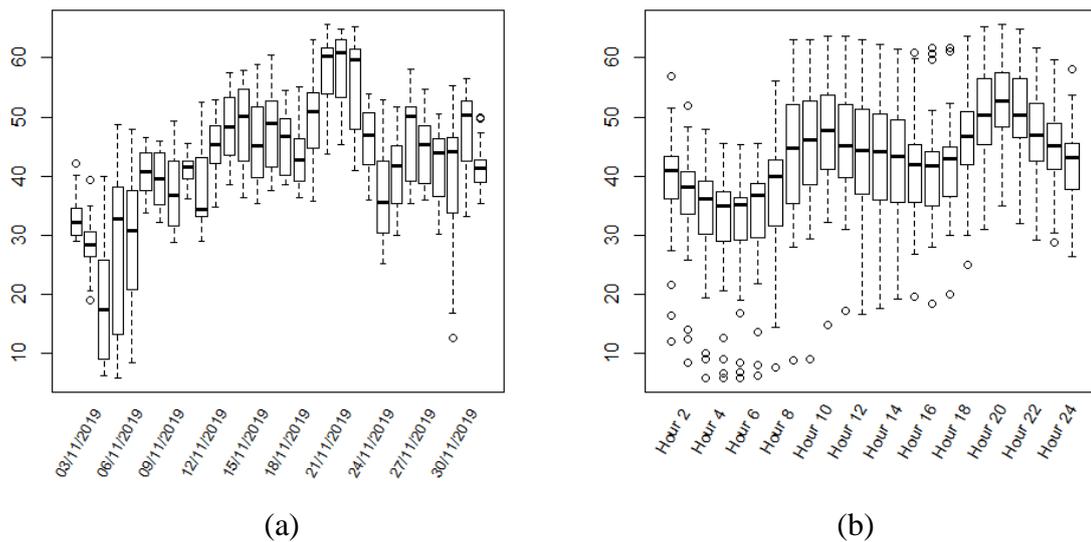

(a) (b)

**Figure 2. Boxplot of the hourly prices in €/MWh in the day-ahead market of Spain in Nov, 2019 in daily (a) and hourly (b) basis.**

Figure 3 presents an example of the market prices on 28[th] November, 2019. Due to the daily demand pattern, the highest price occurs at Hour 20 in the evening and the lowest price occurs on the Hour 4 in the early morning within the day-ahead market. For daily energy arbitrage application within the day-ahead market, batteries should be charged at Hour 4 and discharged at Hour 20.

Batteries may also perform energy arbitrage by providing services in other markets, due to the flexibility requirement and real-time system status, the price volatility for the ancillary services is generally much higher than the day-ahead market. For example, in the same day, the lowest available price would be the Deviation management service from Hour 3 to Hour 5. It is observed that the ancillary services and imbalance costs may significantly increase the price volatility that battery storage system may receive.

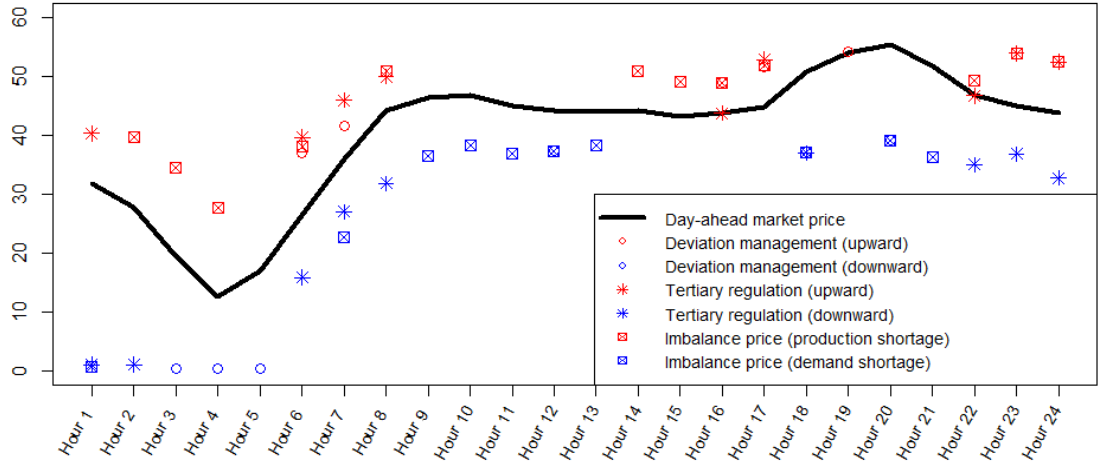

**Figure 3. Market price for Day-ahead market, Deviation management, Tertiary regulation, and Imbalance costs in €/MWh on 28th November, 2019.**

Table 2 presents the mean, median, and standard deviation of the market price of the day-ahead market, related ancillary services, and imbalance cost. It also shows the mean of the daily standard deviation of the day-ahead market price. Although the mean and median for these markets and services are similar, the price variation for the ancillary services and imbalance cost are significantly higher than the day-ahead market.

The daily profit is calculated assuming that the battery storage will be charged at the lowest price of the day and discharged at the highest price. An 85% round-trip efficiency is considered and the energy loss is equally distributed along the charging and discharging phases. In case the calculated profit is less than zero (mainly due to the energy loss in the charging/discharging process), we consider the final profit for that specific operating day is zero (In this case the battery operator would choose not to perform energy arbitrage on that specific day). Note that this is an ideal case, not achievable in practice.

**Table 2. Mean, median, and standard deviation of the prices for the Day-ahead market, Deviation management, Tertiary reserve and Imbalance cost, in €/MWh. (2015-2019)**

|  | Day-ahead price | | | Deviation management | | | Tertiary reserve | | | Imbalance cost | | |
| --- | --- | --- | --- | --- | --- | --- | --- | --- | --- | --- | --- | --- |
| Year | Mean | Median | Std. | Mean | Median | Std. | Mean | Median | Std. | Mean | Median | Std. |
| 2015 | 50.32 | 51.20 | 12.37 | 52.32 | 53.26 | 16.32 | 49.00 | 51.97 | 20.36 | 49.18 | 50.29 | 18.10 |
| 2016 | 39.67 | 40.20 | 14.90 | 40.39 | 40.32 | 16.89 | 37.71 | 40.03 | 19.72 | 38.79 | 40.20 | 17.89 |
| 2017 | 52.24 | 51.04 | 12.28 | 51.64 | 48.87 | 18.22 | 50.51 | 50.70 | 19.40 | 49.81 | 48.80 | 17.83 |
| 2018 | 57.29 | 60.00 | 12.80 | 60.77 | 62.64 | 13.64 | 55.38 | 58.33 | 18.06 | 56.18 | 58.01 | 16.34 |
| 2019 | 47.68 | 48.95 | 10.88 | 48.37 | 50.97 | 13.80 | 49.02 | 53.05 | 15.56 | 47.19 | 49.18 | 19.32 |

Table 3 presents the simulated average cycle (daily) profit assuming that the battery storage system knows perfectly the price in advance, and the total number of profitable days with simulated profit higher than the battery wear cost (100 €/MWh). The results show that the average cycle profit is generally decreasing from 2015 to 2019. This is mainly due to the deceasing in both market price volatility and daily price variation. Although it is an ideal case, the battery storage will obtain a significant higher cycle profit when participating in the ancillary services and imbalance correction since the price volatility is much higher for these markets. However, if we take a battery wear cost of 100 €/MWh, even the highest theoretical daily profit considering ancillary services and imbalance cost is far from reaching the break-even point. There are only very few days when the profit of energy arbitrage of battery systems would cover the corresponding battery wear cost in the period under study (2015-2019) even with perfect price information. The reason behind this fact is the current low energy price volatility in Spain which suggests that the power generation capacity would generally cover demand in most cases, and price peaks caused by lack of capacity are rare.

**Table 3. Average daily profit and total number of profitable days with perfect price information.**

|  | Market and Services | | | |
|---|---|---|---|---|
|  | Day-ahead market only | | Day ahead market, Ancillary services, and Imbalance cost | |
| Year | Average daily profit in € | Profitable days | Average daily profit in € | Profitable days |
| 2015 | 16.81 | 0 | 41.09 | 1 |
| 2016 | 12.71 | 0 | 32.98 | 2 |
| 2017 | 11.09 | 0 | 34.38 | 1 |
| 2018 | 9.63 | 0 | 28.30 | 0 |
| 2019 | 9.42 | 0 | 27.57 | 1 |

### 4.2. Secondary reserve

In the secondary reserve market, the flexibility requirement is higher than the energy based markets, and the required response time would be less than 100 seconds. When participating in the secondary reserve, in addition to the net energy generation or consumption during the service, the service provider will also receive remuneration in terms of the power capacity band. The secondary reserve is an hourly product for which the auction is exercised the day before delivery day and after the day-ahead market and the management of technical restrictions. The payment for power capacity band can be calculated as the product of the contracted power capacity and its band price, a the payment of net energy is calculated as the product of the net energy deviation with respect to the original program and the energy price of secondary regulation of the hour.

Table 4 presents the market information for the secondary reserve market in Spain from 2015 through 2019. The average power band price decreases from 19.57 €/MW to 8.31 €/MW in 2019. The total assignment of secondary reserve band also decreases for the upward band. In order to evaluate the energy usage from secondary regulation, an average energy use rate per unit of power band is assumed, and the *Average Band Utilization* is calculated as the total energy used divided by the total power band assigned. It is observed that the *Average Band Utilization* has been decreasing for upward secondary reserve during the time period, and increasing for the downward power band.

**Table 4. Power band price, total assigned band, total energy usage, and average band utilization for the secondary reserve market in Spain. (2015-2019)**

| Year | Band Price in €/MW | Total Upward Band Assigned in MW | Total Upward Energy Used in MWh | Average upward band utilization MWh/MW | Total Downward band Assigned in MW | Total Downward Energy Used in MW | Average downward band utilization MWh/MW |
|---|---|---|---|---|---|---|---|
| 2015 | 19.58 | 6002468 | 1366302 | 0.23 | -4477793 | -1193013 | 0.27 |
| 2016 | 15.56 | 5989670 | 1529974 | 0.26 | -4468333 | -1012330 | 0.23 |
| 2017 | 14.26 | 5970916 | 1203337 | 0.20 | -4498964 | -1206475 | 0.27 |
| 2018 | 12.56 | 5400159 | 1086235 | 0.20 | -4519135 | -1506230 | 0.33 |
| 2019 | 8.31 | 5203169 | 970742 | 0.19 | -4352156 | -1678825 | 0.39 |

The final payments of the secondary reserve are made according to the provided power capacity band (€/MW) and net energy effectively delivered (€/MWh). Table 5 presents the average day-ahead market price, the average energy price for secondary regulation. It also shows a calculated *Average Profit/Cost of Energy*, which is considered as the cost for energy delivered in secondary reserve and balanced at the average day-ahead market price.

$$Average\ Profit\ /\ Cost\ of\ Energy = \begin{cases} \varepsilon_1 * P_u - \frac{P_{da}}{\varepsilon_2}, \text{(upward regulation)} \\ \varepsilon_1 * P_{da} - \frac{P_d}{\varepsilon_2}, \text{(downward regulation)} \end{cases},$$

Where,
$P_u$ is the average energy price of the upward secondary regulation
$P_d$ is the average energy price of the downward secondary regulation
$P_{da}$ is the average energy price of the day-ahead market
$\varepsilon_1$ is the efficiency for battery discharging
$\varepsilon_2$ is the efficiency for battery charging

In this study, an 85% round-trip efficiency is considered for the battery storage and the energy loss is evenly distributed between charging and discharging phases. As a result, the efficiencies can be calculated as,

$$\varepsilon_1 = \varepsilon_2$$
$$\varepsilon_1 * \varepsilon_2 = 85\%$$

Considering the sum of the battery wear cost and profit/cost in energy terms as the effective energy usage cost, the *Break-Even Power Band Price* can be calculated as the *Effective Energy Usage Cost* divided by the *Average Band Utilization* of the secondary band.

In other words, it is only profitable for battery storage systems to provide secondary reserve when the remuneration from the power band is higher than the sum of the battery wear cost and the energy profit/cost. And the *Effective Energy Usage Cost* can be calculated as the sum of the *Battery Wear Cost* and the *Average Profit/Cost of Energy*.

**Table 5. Average price of the day-ahead market and secondary regulation, and average profit/cost of delivering energy from battery storage considering the average band utilization. (2015-2019)**

|  | Average day-ahead market price in €/MWh | Average energy price of secondary regulation in €/MWh | | Average profit/cost of delivering energy from battery storage in €/MWh | |
|---|---|---|---|---|---|
| Year |  | Upward | Downward | Upward | Downward |
| 2015 | 50.32 | 53.71 | 40.11 | -5.29 | 2.71 |
| 2016 | 39.67 | 44.09 | 33.21 | -2.55 | 0.40 |
| 2017 | 52.24 | 54.60 | 45.05 | -6.55 | -0.90 |
| 2018 | 57.29 | 58.05 | 50.04 | -8.86 | -1.68 |
| 2019 | 47.68 | 51.25 | 40.63 | -4.67 | -0.30 |

Table 6 presents the calculated *Break-Even Power Band Price* and the total number of profitable hours with power band price higher than the break-even price. Given the battery wear cost of today, in 2015, a battery storage system would be theoretically profitable during more than 2000 hours providing secondary reserve in both directions. However, due to the significant fall in power band price, in 2019 this number decreased to around 600 hours in upward reserve and 74 in downward reserve.

Still, we would like to emphasize the fact that thanks to the payment of flexibility, this number of hours that battery storage is theoretically profitable is much higher than the one resulting from the application of energy arbitrage. In addition, the results show that the competition of flexible resources is fierce in the Spanish electricity market and battery storage technics with the current cost may start joining the competition.

**Table 6. Estimated break-even band price and the total number of profitable hours. (2015-2019)**

|  | Break-even power band price in €/MW | | Total number of profitable hours | |
|---|---|---|---|---|
| Year | Upward | Downward | Upward | Downward |
| 2015 | 23.97 | 25.92 | 2347 | 1880 |
| 2016 | 26.20 | 22.57 | 1135 | 1609 |
| 2017 | 21.47 | 27.06 | 1272 | 691 |
| 2018 | 21.90 | 33.89 | 759 | 237 |
| 2019 | 19.53 | 38.69 | 596 | 74 |

### 4.3. Sensitivity analysis

Next, the explanation of how the decline in battery cost would affect the market potential using the proposed potentially profitable utilization time analysis is given. Rather than modeling the battery wear cost as a single value as demonstrated in the previous sections, sensitivity analysis is conducted for the cost range from zero to the empirically estimated battery wear cost.

Figure 4 presents the potentially profitable utilization time for applying a battery system for energy arbitration within only the day-ahead market in Spain in 2019, given that the battery wear cost varies from 100 €/MWh to 0 €/MWh. It is shown that the potentially profitable utilization time remains near zero if the battery wear cost is above 40 €/MWh.

Then the potentially profitable utilization starts to increase with a slow-moderate rate in the cost range from 40 €/MWh to 20 €/MWh. After that, the potential utilization increases rapidly if the battery cost is below 20 €/MWh. Finally, if there is no battery cost, a battery system could be profitable for energy arbitrage applications (only within the day-ahead market) on approximately 310 days in 2019. For the rest of the days, the day-ahead market price difference is just not large enough to make it profitable considering the energy losses in the charging-discharging cycle.

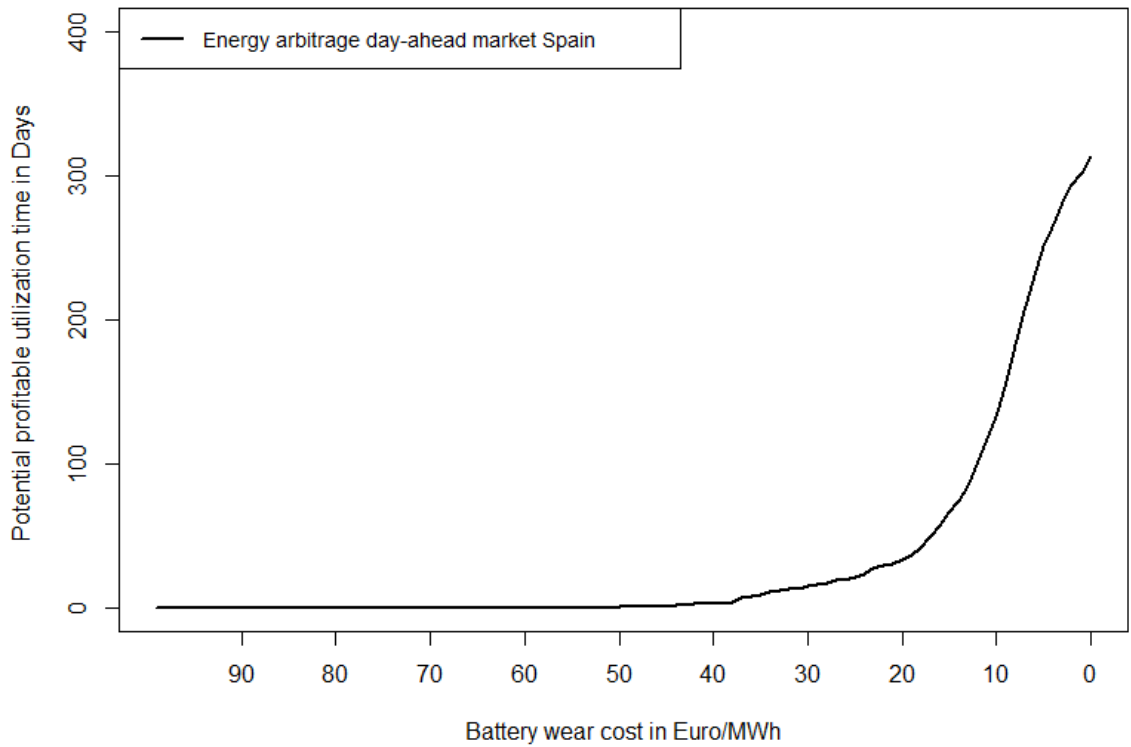

**Figure 4. Potentially profitable utilization time (in Days) for energy arbitrage application within the day-ahead market in Spain, 2019.**

The potentially profitable utilization time curve also makes it comparable for market applications or services with different remuneration structures or in different countries. Figure 5 presents the normalized potentially profitable utilization time for applications of energy arbitrage within the day-ahead market, energy arbitrage with ancillary services and the secondary reserve in Spain during 2019. Meanwhile, in order to demonstrate the application of the presented method in an international context, we also present the normalized potentially profitable utilization time for the energy arbitrage application within the French day-ahead market in 2019 in Figure 5.

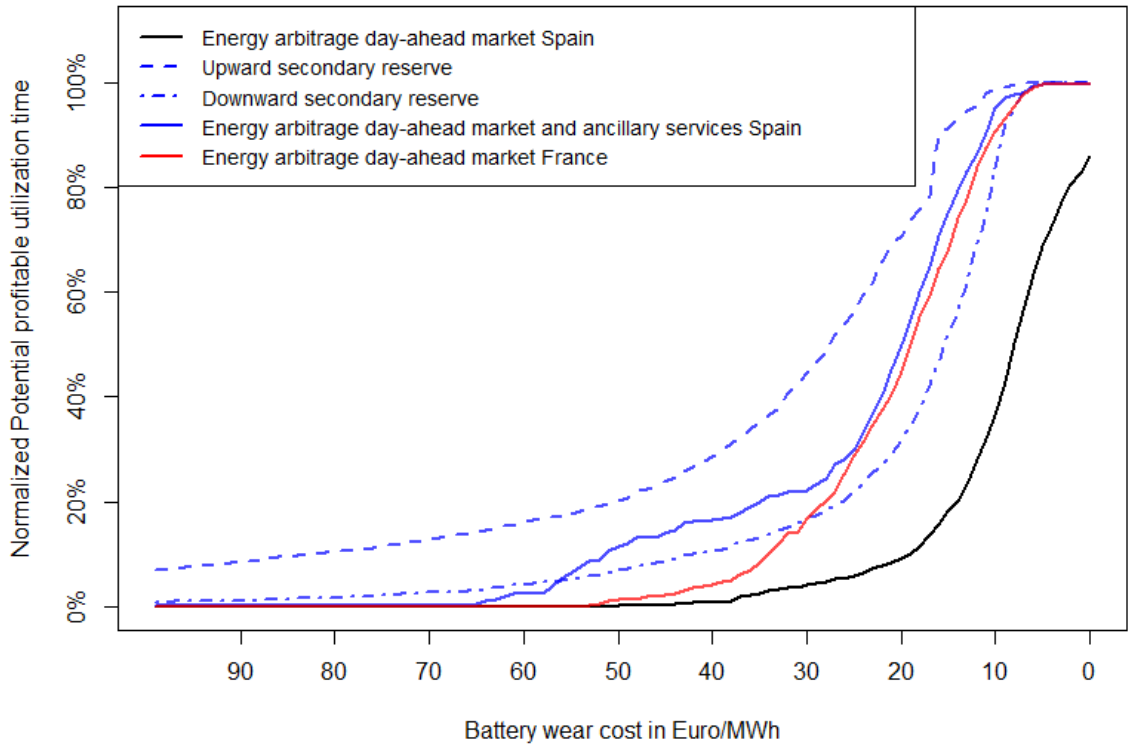

**Figure 5. Normalized potentially profitable utilization time for grid applications in Spain and energy arbitrage within the day-ahead market in France, 2019.**

Among the grid-connected applications in the Spanish electricity market, the upward secondary reserve is the most promising application for battery systems, and its rate of potentially profitable utilization time is higher than any other applications for the whole range of battery cost. The potentially profitable utilization rate would be more than 20% of the total time, and increases rapidly if the battery cost falls below 50 €/MWh. On the other hand, the downward secondary reserve is not as profitable as the upward one, the potential utilization may reach 20% only when the battery wear cost is below 28 €/MWh. The reason behind this difference is mainly due to the fact that the system operator generally requires less power band for the downward secondary reserve, and the *Average Band Utilization* for providing the downward secondary reserve is higher. As a result, it would require a higher capacity payment to balance the higher battery usage.

It should be also noted that the potential profitability of providing both upward and downward secondary reserve can be improved by combining these two applications and providing these two services at the same time. Since the energy used in providing secondary reserve for one direction can be rebalanced when the system operator requires energy for the other direction in the same hour. Due to the fact that the model combining secondary reserve in both directions would involve complex optimization model, we would leave it as a possible future research path.

The potentially profitable time for energy arbitrage within the day-ahead market and ancillary services (tertiary reserve, deviation management and imbalance cost) lies between the upward and downward secondary reserves when the battery wear cost is below 57 €/MWh. And the potentially profitable utilization time would rise if the battery wear cost is below 30 €/MWh. Finally, the energy arbitrage application within only the day-ahead market would be the less promising application which requires a battery cost below 15 €/MWh for reaching a 20% of potentially profitable utilization rate.

When comparing the Spanish and the French market, the energy arbitrage application in the French day-ahead market clearly shows a much higher potentially profitable utilization rate. Indeed, the day-ahead prices in France are more volatile due to the following reasons: On the one hand, in the French market, a large part of the energy is supplied by energy plants with low flexibility such as the nuclear power plants, and the main flexibility provider, such as oil and gas turbines and hydro power stations, would charge a much higher price for the peak hours. While in Spain, the combined cycle gas turbine plants with high flexibility represent for almost 30% of the total installed capacity. As a result, the competition of the flexibility market in Spain is more intense. On the other hand, the market biddings in the day-ahead market are limited from 0 €/MWh to 180.3 €/MWh (Those limits are highly likely to be removed in the future). While in France, there is no such limit, so that the prices in France could be negative or reach a much higher level than in Spain.

The result mentioned above reveals the fact that there is a positive correlation between the potential profitability and the flexibility requirement of the market/service. Among the applications in the Spanish market, the secondary reserve requires the highest flexibility (hundreds of seconds), followed by the ancillary services such as the tertiary reserve (about 15 minutes) and the deviation management service (about one hour). And the imbalance correction requires that the battery is able to respond to the power imbalance within one hour. For these applications, the potential profitability is clearly higher than the day-ahead applications which are programed in at least 12 hours ahead. And from an international point of view, the comparison between the Spanish and the French day-ahead market also shows that grid-connected battery storage systems have a higher potential in electricity markets where higher flexibility is demanded.

# 5. Discussions and conclusion

## 5.1. Other related issues in Spain

Apart from the market profitability, as an emerging technology, battery storage for grid-connected applications also faces regulatory and market design barriers.

### 5.1.1. Technological definition and recognition

Theoretically, battery systems may work as a standalone unit in the electricity market and provide ancillary services such as the secondary reserve services, similar to the existing pumped hydro plants.

However, one of the important concerns nowadays when developing battery storage projects in Spain would be the lack of related supporting regulations. A new standalone battery storage project may need a permit from the system operator REE if the energy storage system is connected to the grid directly. However, there are no regulations related to battery systems, so a battery storage system does not belong to any existing types of power generating or consuming units.

Currently in Spain, almost all the pilot battery systems, that have been installed so far, have been integrated in an existing generation facility, and in most cases, the integration only requires industrial permits, but not REE permits (since the device does not increase the evacuation power capacity and batteries are considered as behind-the-meter storage devices).

This kind of integration has its drawback. When a battery system is integrated into an existing power plant, the available power capacity of the battery is limited to the power capacity of the plant. Also, the integrated power plant is normally registered as a power generating unit. In this case, the battery is able to absorb energy only from the power plant itself but not from the grid. Moreover, for some special applications such as the secondary reserve, the power deviation from the power plant may cause additional problems since balancing the power deviation with the battery is a prerequisite for providing secondary reserve.

### 5.1.2. Market design

The lack of recognition of the battery technologies directly implies that the market is not designed to accommodate and encourage a wide deployment of battery storage systems. The electricity grid in Spain has generally relied on dispatchable thermal power plants and hydro power units for supporting system security and providing ancillary services [43]. Therefore, market design is considered to be a centralized regulation solution based on large energy groups or large consumers. Power flexibility products are commonly defined using the terminology of thermal power plants and only allow the participation of conventional dispatchable power plants.

For example, the technical constrains, including those of for transmission and distribution, are currently solved by thermal power plants. The system operator pays the thermal plant to run at its minimum power capacity in some specific areas where technical constrains exist.

In a similar way to the technical constrains, the peak load problem is solved only by thermal power plant with the service of additional upward power reserve, in such a way that the participating generation unit is paid to schedule the minimum capacity in the adjusting market.

For secondary reserve markets, the service is offered and liquidated by regulating zones. The regulating zones in Spain are the large energy groups such as Iberdrola and Endesa. Since the secondary regulation only requires a minimum respond rate, the participating power plants may response to the order with different speed. This may cause additional difficulties in allocating incomes and costs within the regulating zone, especially in case some of the power plants in the same regulating zone do not fulfill the secondary regulation requirements.

On the other hand, the secondary reserve auction only opens once a day after the day-ahead market. This would not cause serious problems for large conventional power plants and/or groups of power plants. However, for distributed energy sources with relatively low power and energy capacity, real-time adjustment measures would be essential.

Compared with other European countries, the power capacity from hydro power plants is relatively high and provides a significant part of the ancillary services in Spain. Although different research works have explored the profitability of battery storage systems in primary reserves, it is not expected in the near future that Spain will join the central European primary reserve market or change the non-remunerable structure of the primary reserve, since the cost of providing frequency support from hydro power is relatively low.

## 5.2. Conclusion

Battery storage systems are expected to be an essential part in the future energy system due to their flexible and distributed nature. In this study, the major barriers to a wide deployment of grid-connected battery storage systems in Spain are investigated. Although the cost of battery systems has been reduced in the recent years, the battery cycle lifetime compared with the installation cost is still the main technological barrier for grid-connected applications. In this study, the results show that the battery wear cost resulting from degradation is still too high compared with the marginal generation cost of other technologies, and this would prevent battery systems from being profitable in energy-based markets and energy arbitrage applications. For example, it requires a battery wear cost below 15 €/MWh for the application of energy arbitrage within the day-ahead market to reach a 20% potentially profitable utilization rate.

With the proposed utilization analysis framework, the future potential of the market profitability can be compared for given battery cost reduction. Market profitability analysis also claims that under a deregulated electricity market such as the Spanish one, although it might be not completely economically feasible nowadays, higher potential is clearly shown for grid-connected battery systems when higher flexibility is required. When ancillary services are considered, the minimum battery wear cost required for 20% potentially profitable utilization rate increases to 35 €/MWh. And this value can be still improved to 50 €/MWh for providing upward secondary reserve. The results also show that for the same application purpose, battery storage system would obtain a higher potential utilization rate in the countries where higher flexibility is demanded.

Currently in Spain, both the regulatory and market frameworks are designed based on a centralized power system supported by large power plants and energy groups. As a novel technology, battery storage systems are not able to be operated independently in the market. Again, the lack of awareness and recognition of the technology directly leads to missing of supporting regulations and market products that would address the comparative advantages of battery systems.

To summarize, the numerical and empirical analysis suggests that the high cycle cost for batteries is still the main barrier for grid-connected battery systems, and the flexibility offered by such systems would be currently the most promising comparative advantage for this novel technology. Besides providing directly subsidies to the development of battery projects, a correct recognition of the barriers and advantages by all the stakeholders, including the system/market operator, policy maker, investor, and project manager, is also a key factor to promote battery storage technologies in grid-connected applications.

This paper focuses on the grid applications for battery systems under the Spanish electricity market. Indeed, with the help of the proposed framework, more research could be conducted for a wider range of countries, which requires the special knowledge of the energy market and the ancillary services in each country. Meanwhile, other common grid applications, which are not currently remunerable in Spain, such frequency regulation (especially primary regulation), system capacity reduction, peak shaving, etc. should be elaborated with more detail in the future under their market designs.

Future research work should also focus on, technologically, reducing the energy equivalent cycle cost for battery systems, and economically, exploring new market design and valuation model that address the flexibility provided by the battery systems. For market and system regulators, more efforts are necessary to reduce the barriers from regulatory framework, promote pilot projects, and design appropriate market products or services that adequately address the flexibility provided by different technologies.

# References

[1] Schmidt, O., Hawkes, A., Gambhir, A., and Staffell, I. (2017). The future cost of electrical energy storage based on experience rates. Nature Energy, 2(8), 17110.

[2] IRENA (2017), Electricity Storage and Renewables: Costs and Markets to 2030, International Renewable Energy Agency, Abu Dhabi.

[3] Tsiropoulos, I., Tarvydas, D., & Lebedeva, N. (2018). Li-ion Batteries for Mobility and Stationary Storage Applications Scenarios for Costs and Market Growth. Publications Office of the European Union: Luxembourg.

[4] IRENA (2019), Innovation landscape brief: Utility-scale batteries, International Renewable Energy Agency, Abu Dhabi.

[5] Wang, N., & Mogi, G. (2017). Deregulation, market competition, and innovation of utilities: Evidence from Japanese electric sector. Energy Policy, 111, 403-413.

[6] Pollitt, M. G. (2019). The European single market in electricity: an economic assessment. Review of Industrial Organization, 55(1), 63-87.

[7] Mayer, K., & Trück, S. (2018). Electricity markets around the world. Journal of Commodity Markets, 9, 77-100.

[8] Ding, Y., Shao, C., Yan, J., Song, Y., Zhang, C., & Guo, C. (2018). Economical flexibility options for integrating fluctuating wind energy in power systems: The case of China. Applied Energy, 228, 426-436.

[9] Yan, J., Lai, F., Liu, Y., David, C. Y., Yi, W., & Yan, J. (2019). Multi-stage transport and logistic optimization for the mobilized and distributed battery. Energy conversion and management, 196, 261-276.

[10] Jülch, V. (2016). Comparison of electricity storage options using levelized cost of storage (LCOS) method. Applied energy, 183, 1594-1606.

[11]Chawla, M., Naik, R., Burra, R., and Wiegman, H. (2010). Utility energy storage life degradation estimation method. In 2010 IEEE Conference on Innovative Technologies for an Efficient and Reliable Electricity Supply (pp. 302-308). IEEE.

[12] Xu, B., Oudalov, A., Ulbig, A., Andersson, G., and Kirschen, D. S. (2016). Modeling of lithium-ion battery degradation for cell life assessment. IEEE Transactions on Smart Grid, 9(2), 1131-1140.

[13] Zhou, C., Qian, K., Allan, M., and Zhou, W. (2011). Modeling of the cost of EV battery wear due to V2G application in power systems. IEEE Transactions on Energy Conversion, 26(4), 1041-1050.

[14] Ashari, M., Nayar, C. V., and Keerthipala, W. W. L. (2001). Optimum operation strategy and economic analysis of a photovoltaic-diesel-battery-mains hybrid uninterruptible power supply. Renewable energy, 22(1-3), 247-254.

[15] Zubi, G., Dufo-López, R., Carvalho, M., and Pasaoglu, G. (2018). The lithium-ion battery: State of the art and future perspectives. Renewable and Sustainable Energy Reviews, 89, 292-308.

[16] TESLA (2017) Tesla Powerwall limited warranty (USA), effective date: April 19, 2017. https://www.tesla.com/sites/default/files/pdfs/powerwall/powerwall_2_ac_warranty_us_1-4.pdf. (Cited on 30 Jan, 2020)

[17] Teng, F., Pudjianto, D., Strbac, G., Brandon, N., Thomson, A., and Miles, J. (2015). Potential value of energy storage in the UK electricity system.

[18] Gholinejad, H. R., Loni, A., Adabi, J., & Marzband, M. (2020). A hierarchical energy management system for multiple home energy hubs in neighborhood grids. Journal of Building Engineering, 28, 101028.

[19] Marzband, M., Azarinejadian, F., Savaghebi, M., Pouresmaeil, E., Guerrero, J. M., & Lightbody, G. (2018). Smart transactive energy framework in grid-connected multiple home microgrids under independent and coalition operations. Renewable energy, 126, 95-106.

[20] Nayak, C. K., Kasturi, K., & Nayak, M. R. (2019). Economical management of microgrid for optimal participation in electricity market. Journal of Energy Storage, 21, 657-664.

[21] Xu, X., Hu, W., Cao, D., Huang, Q., Liu, Z., Liu, W., ... & Blaabjerg, F. (2020). Scheduling of wind-battery hybrid system in the electricity market using distributionally robust optimization. Renewable Energy.

[22] Zhang, Y., Ma, T., Campana, P. E., Yamaguchi, Y., & Dai, Y. (2020). A techno-economic sizing method for grid-connected household photovoltaic battery systems. Applied Energy, 269, 115106.

[23] Núñez-Reyes, A., Rodríguez, D. M., Alba, C. B., & Carlini, M. Á. R. (2017). Optimal scheduling of grid-connected PV plants with energy storage for integration in the electricity market. Solar Energy, 144, 502-516.

[24] Faraji, J., Ketabi, A., & Hashemi-Dezaki, H. (2020). Optimization of the scheduling and operation of prosumers considering the loss of life costs of battery storage systems. Journal of Energy Storage, 31, 101655.

[25] Kusakana, K. (2020). Optimal peer-to-peer energy management between grid-connected prosumers with battery storage and photovoltaic systems. Journal of Energy Storage, 32, 101717.

[26] Paliwal, P. (2020). Reliability constrained planning and sensitivity analysis for Solar-Wind-Battery based Isolated Power System. International Journal of Sustainable Energy Planning and Management, 29, 109-126.

[27] Lüth, A., Zepter, J. M., del Granado, P. C., & Egging, R. (2018). Local electricity market designs for peer-to-peer trading: The role of battery flexibility. Applied energy, 229, 1233-1243.

[28] Mirtaheri, H., Bortoletto, A., Fantino, M., Mazza, A., & Marzband, M. (2019, June). Optimal Planning and Operation Scheduling of Battery Storage Units in Distribution Systems. In 2019 IEEE Milan PowerTech (pp. 1-6). IEEE.

[29] Olk, C., Sauer, D. U., & Merten, M. (2019). Bidding strategy for a battery storage in the German secondary balancing power market. Journal of Energy Storage, 21, 787-800.

[30] Stroe, D. I., Knap, V., Swierczynski, M., Stroe, A. I., and Teodorescu, R. (2016). Operation of a grid-connected lithium-ion battery energy storage system for primary frequency regulation: A battery lifetime perspective. IEEE Transactions on Industry Applications, 53(1), 430-438.

[31] Zhu, D., and Zhang, Y. J. A. (2018). Optimal coordinated control of multiple battery energy storage systems for primary frequency regulation. IEEE Transactions on Power Systems, 34(1), 555-565.

[32] Fiorini, L., Pagani, G. A., Pelacchi, P., Poli, D., and Aiello, M. (2017). Sizing and siting of large-scale batteries in transmission grids to optimize the use of renewables. IEEE Journal on Emerging and Selected Topics in Circuits and Systems, 7(2), 285-294.

[33] Sun, Y. K., Ding, Q., Xing, H., Zeng, P. P., and Sun, F. F. (2018, November). Optimal Placement and Sizing of Grid-scale Energy Storage in Distribution Networks with Security Constrains. In 2018 International Conference on Power System Technology (POWERCON) (pp. 1477-1482). IEEE.

[34] Koller, M., Borsche, T., Ulbig, A., and Andersson, G. (2015). Review of grid applications with the Zurich 1 MW battery energy storage system. Electric Power Systems Research, 120, 128-135.

[35] Skoglund, A., Leijon, M., Rehn, A., Lindahl, M., & Waters, R. (2010). On the physics of power, energy and economics of renewable electric energy sources-Part II. Renewable energy, 35(8), 1735-1740.

[36] Zhixin, W., Chuanwen, J., Qian, A., & Chengmin, W. (2009). The key technology of offshore wind farm and its new development in China. Renewable and Sustainable Energy Reviews, 13(1), 216-222.

[37] OMIE (2018): Day-ahead and Intraday Electricity Market Operating Rules, May 2018. http://www.omie.es/files/market_rules_2018.pdf. (Cited on 30 Jan, 2020)

[38] Bueno-Lorenzo, M.; Moreno, M. A.; Usaola, J. (2013). Analysis of the imbalance price scheme in the Spanish electricity market: A wind power test case. Energy policy, 62: 1010-1019.

[39] Lobato, M., Egido, C., Rouco, R., & Lopez, C. (2008). An overview of ancillary services in Spain. Electric Power Systems Research, 78(3), 515-523.

[40] UCTE (2009), UCTE Operation Handbook – Policy 1: Load-Frequency Control and Performance. Technical Report UCTE OH P1, Union for the Co-ordination of Transmission of Electricity.

[41] De la Fuente, I. (2009). Ancillary Services in Spain: dealing with High Penetration of RES. Red Eléctrica, Madrid, 210.

[42] Zafirakis, D., Chalvatzis, K. J., Baiocchi, G., and Daskalakis, G. (2016). The value of arbitrage for energy storage: Evidence from European electricity markets. Applied energy, 184, 971-986.

[43] Fernandes, C., Frías, P., & Reneses, J. (2016). Participation of intermittent renewable generators in balancing mechanisms: A closer look into the Spanish market design. Renewable Energy, 89, 305-316.